# Multiple Superconducting Transitions in α-Sn Films Grown by Molecular Beam Epitaxy

Yuanfeng Ding, Huanhuan Song, Jinshan Yao, Lian Wei, Junwei Huang, Hongtao Yuan, and Hong Lu

*College of Engineering and Applied Sciences, Nanjing University*
*National Laboratory of Solid State Microstructures, Nanjing University*

Gray tin, also known as α-Sn, has been attracting research interest recent years due to its topological nontrivial properties predicted theoretically. The Dirac linear band dispersion has been proved experimentally by angle resolved photoemission spectroscopy. We have grown a series of α-Sn thin film samples in two types with different substrates and thicknesses by molecular beam epitaxy. To explore the possible exotic physical properties related to the topological band structures, we have measured the electrical transport properties of our α-Sn thin film samples and observed multiple superconducting transitions. We have identified the transitions above 4.5 K, besides the transition maybe related to the β phase around 3.7 K. The changes of the superconducting properties over time reflect the aging effects in our samples. We have also confirmed the strain effects on the superconducting transitions through altering the relative thickness of our samples.

**Introduction**

The electrical and magnetic properties of α-Sn have been extensively investigated since 1950s as it was found to be a semiconductor similar to silicon and germanium.[1] There were arguments about the accuracy of the results until the single crystalline α-Sn in the form of filaments and bulk was prepared by Ewald and coworkers.[2,3] However, almost all the measurements were conducted at low temperatures at that time[2,4-6] since the phase transition from α to β occurs spontaneously around 13.2 °C. Recent years, people can fabricate high-quality semiconductor thin films by molecular beam epitaxy (MBE) or metal-organic chemical vapor deposition (MOCVD). Farrow et al. firstly deposited α-Sn films on InSb and CdTe substrate at room temperature by MBE and found the transition temperature was raised to 70 °C.[7] The α-Sn thin films can exist at higher temperature under the strain imposed by the substrate due to the small lattice mismatch between the α-Sn and InSb or CdTe.[7-9]

On the other hand, following the development of topological physics, emerging materials with novel quantum phenomena have been found, and α-Sn under strain was predicted to be a member of them.[10-14] Dependent on the strain applied, α-Sn can be transformed into topological insulator (TI) or Dirac semimetal (DS).[10,13-15] While the linear band dispersion and Dirac cones have been observed by angle resolved photoemission spectroscopy (ARPES),[15-18] however, few related electrical transport measurement has been reported as those on $Cd_3As_2$ and $Na_3Bi$.[19-21] Since the most α-Sn films were grown on relatively conducting InSb substrates, the electrical measurement becomes difficult. Nevertheless, the superconducting properties can be confirmed and investigated as long as the substrate is non-superconducting in the temperature range of interest. Though the β-Sn is one of the earliest identified superconducting metals with critical temperature about 3.72 K, it is claimed that the α-Sn is not superconductive, at least in the bulk form. Fortunately, quantum size effects or confinement effects under low dimension significantly change the material properties. Stanene, single atomic layer of α-Sn, has been predicted to hold topological superconductivity[21,22] and successfully grown on $Bi_2Te_3$(111) by Zhu et al..[23] Superconductivity has been observed by Xue et al. in few-layer stanene, grown on PbTe/Bi2Te3/Si(111).[24] They investigated the thickness dependence of the superconducting properties within 20 bilayers, namely no more than 10 nm. Whether the superconductivity of stanene will retain or how the critical behavior will change with larger thickness remains questions.

In this work, we report superconducting transitions in α-Sn films several tens to hundreds nanometers thick grown on InSb with two type templates by MBE. Besides the β-like transition around 3.7 K, we also observed extra transitions above 4.5 K. From the investigation on the aging effects and strain effects, we discuss the possible sources of these superconducting phases.

**Method**

The α-Sn samples used in this research were grown on InSb(001) or GaAs(001) substrates by MBE, as shown in Figure 1(a), named type A and type B, respectively. Type B samples were designed to reduce the contribution of the substrate in the electrical measurement and facilitate the modulation of the strain in the epitaxial films. Some of type B samples were capped with amorphous Si layers to avoid oxidation. Figure 1(b) shows the X-ray diffraction (XRD) patterns of two type samples. The α-Sn peaks can be clearly seen. The fringes suggest the high crystalline quality and smooth surfaces and interfaces of type A samples. All the samples is fully strained. The details about the growth and structural characterization are presented elsewhere. Considering the stable phase at room temperature is usually the β phase due to the relatively low phase transition temperature, we carefully checked the thermal stability of our samples. Figure 1(c) shows the phase transition temperatures of type A samples with different thicknesses estimated by temperature dependent XRD and Raman spectra. Note that the lowest phase transitions temperature is still above 70 °C for the thickest sample of 400 nm. The raised phase transition temperature is usually attributed to the stabilization effect of the substrate, and we also calculated the out-of-plane residual strain(?) as a function of the thickness of α-Sn films using plastic flow model (?), showing the same tendency as the phase transition temperature.

Transport measurements on our α-Sn samples were conducted in a Quantum Design physical property measurement system (PPMS) and an Oxford Instrument TeslatronPT system with temperature as low as 1.5 K and magnetic field up to 14 T. Freshly cut indium particles were cold pressed on to 5 mm × 5 mm square samples cleaved from the quarter of the 2 inch wafer in four-probe configuration. The resistance was measured by both dc method and standard lock-in techniques. Measurements with different orientations relative to the magnetic field can be realized with the horizontal rotator option of PPMS.

**Results**

Superconductivity was initially observed in the measurement on type A α-Sn samples as can be seen in Figure 1(c). While a globally superconducting transition can be seen in the 100 nm sample around 3.7 K, the critical temperature of β phase, a more detailed measurement on the 400 nm sample shows two separate transitions with smaller volume fractions. The other sharp transition appears at 4.96 K, which is obviously above the transition temperature corresponding to β-Sn. Multiple superconducting transitions can also be observed in type B samples as shown in Figure 2 and Figure 3. Figure 3(a) is the resistance curve for a Si capped 100 nm type B α-Sn samples and presents three superconducting transitions including an even higher critical temperature around 6 K. Figure 2 demonstrates the aging effects on the superconducting behavior. To confirm the stability of the superconducting transitions, we measured our samples repeatedly with intervals of tens of days. The sample was placed directly in the air in a dry box at room temperature between measurements. In Figure 2(a), for example, the resistance curve of the 400 nm type A α-Sn sample shows that the transition amplitude around 4.96 K changed obviously over time while the transition corresponding to β-Sn has little change. This observation may indicate that the β phase was formed during growth due to the imperfection of the InSb surface. Figure 2(b) shows another aspect of the aging effects. Measurements on 50 nm type B samples without Si capping shows a change in transition width. Again there is another transition at higher temperature. This gradual transition evolved into a transition with a much smaller width around 4.8 K. The aging effects also exist in other samples with multiple transitions (not shown here) and the higher-$T_c$ transition seems to be more sensitive.

The critical fields for 100 nm and 50 nm type B samples measured at different temperatures and orientations are shown in Figure 4 for example. There are three clear critical fields for the 100 nm sample below 4 K as can be seen in Figure 4(a), suggesting larger critical fields for higher-Tc transitions. The critical fields decrease with increasing temperature. One transition on the curves disappears when the temperature raises up through 4 K, 5 K and 6 K, in well consistent with the three superconducting transitions in Figure 3(a). By fitting of the experimental results to the common relation between the critical field and temperature

$$B_c(T) = B_c(0)\left[1 - \left(\frac{T}{T_c}\right)^2\right]$$

We can obtain the critical field at absolute zero temperature $B_c(0)$ of the 100 nm type B sample is about 1360 Oe and the fitting critical temperature 3.8 K is consistent with experimental values shown in Figure 3(a). In Figure 4(b), the angular dependence of 50 nm sample was measured between 0° and 90° at 2 K. The critical field increases with the angle of the magnetic field relative to the normal of the sample surface as shown in the inset of Figure 4(b) and can be fitted with 2D Tinkham model according to the equation below.[25] The critical field parallel to the sample surface $Bc_{\parallel}$ (90°) is obviously larger than that perpendicular to the surface $Bc_{\perp}$ (0°), indicating 2D character. We note that the critical fields of thinner samples are larger than those of thicker ones

as reported by others.[26]

$$\left(\frac{B_c(\theta)\sin\theta}{B_{c\parallel}}\right)^2 + \left|\frac{B_c(\theta)\cos\theta}{B_{c\perp}}\right| = 1$$

**Discussion**

The superconductivity in α-Sn thin films up to this thickness range has never been reported, though the superconductivity induced by the proximity effect of the β-Sn islands in the films has been investigated before.[26] Note that the critical field decided in Figure 4(a) is actually the critical field for the transition at the lowest $T_c$ (3.94 K) in Figure 3(a). This transition probably corresponds to the β phase in the film as discussed above. The resulting critical field 1360 Oe is well above the bulk value about 300 Oe but within the range decided by Didschuns el al. considering the difference between samples and measurement methods.[26] However, Didschuns et al. concluded that the critical temperature of superconductivity induced by proximity effect is lower than that of bulk β-Sn.[26,27] In our results, the critical temperature is close to 3.7 K if not slightly higher. A very high in-plane critical field was decided by Xue et al. to be about 1.2 T for 3-bilayer stanene.[24] In the inset of Figure 4(b), the global critical fields, or indeed the critical fields for the higher-Tc transition at different angles are demonstrated. The in plane critical field (90°) is about 0.82 Tand this value follows the decreasing trend with increasing thickness compared to Xue's result.

According to the results above, more than one superconducting phases exist in our α-Sn samples besides β-Sn. The other phases have higher critical temperatures with higher critical fields, smaller volume fractions (usually not more than 20%), and are sensitive to the aging effects. We have analyzed the possible sources of these transitions. The oxide may also be superconductive but is less possible in our case since we capped the α-Sn sample with 5 nm Si and measured immediately after growth. It should be easy to observe a complete transition if the surface oxide is superconductive, which is not true in our measurement. The oxidation of the InSb should not be the source. We capped several tens nanometers Sb on the InSb layer when transferred it into group-IV system. The deoxidation was done again after decoating Sb. The more persuasive evidence was given in Figure 5 and discussed below.

It is not likely the Indium used as contacts contribute a source of superconductivity since the four-probe method eliminates the contact resistance. The critical temperature of Indium is 3.4 K and the critical field is about 280 Oe,[28] much lower than the above superconducting parameters. However, we observed a change of the sample surface around In contacts. The mirror-like gray surface changed into white circle region around In contacts after several days. We assume it as In-Sn alloy formed by the interdiffusion of In and Sn atoms . According to the literature, the critical temperature of In-Sn alloy covers the range from 3.4 K to 7.3 K depending on the composition.[29,30] Because the diffusion is a gradual process, the aging effect is plausible. To examine the assumption above, we need to eliminate the influence of In contacts. We used Ag paste instead of In and the results are shown in Figure 3. Figure 3(a) demonstrates the resistance-temperature curves measured with In contacts and Ag paste. In contacts and Ag paste are made on the opposite edges of the same Si capped 100 nm type B sample. Despite a little difference in resulting values, three transitions appear on both curves and match well with each other. This gives a strong evidence that the appearance of the higher-Tc transitions is independent of In contacts. Figure 3(b) shows the corresponding resistance curves under different magnetic fields with Ag paste contacts. The relation between the critical temperature and critical field is essentially the same as that with In contacts. All the transitions shift to lower temperatures with increasing field, again illustrating their superconducting character. The measurement was repeated at low temperature without taking the sample out. From Figure 3(c) and (d), we can see that except for small shifts in the first two days, the superconducting transitions have little change for both type contacts within measurement error over four days. Therefore, either high temperature or oxidation may cause the aging effects through In atom diffusion or microstructure changes but not the source of the observed superconducting transition. In fact, there are much less aging effects for samples with Ag paste contacts at room temperature.

To eliminate the contribution of the substrate, we measured three template samples shown in Figure 5(a). The resistance of template samples is much larger than that deposited with α-Sn films and no superconductivity has been observed. The α-Sn samples still show the superconductivity. However, there are some difference with different thickness InSb layers. We use InSb layers to stabilize the α-Sn structure and believe that it is relaxed on GaSb/GaAs with the thickness under investigation due to the large lattice mismatch between InSb and GaAs. Negligible influence of GaSb and GaAs on the superconducting transitions can also be implied in Figure 2(b). Therefore when the α-Sn film thickness becomes comparable to that of InSb, the stabilization effect reduces and the α-Sn structure cannot retain at room temperature for a long time. This may cause the global transition at 3.9 K in Figure 3(a) indicating a large amount of β phase. We reduced the α-Sn film thickness on different thickness InSb layers to suppress the structure transformation. In the enlarged version in Figure 5(b), we see that the superconducting transition is not complete any more. The transition around 5 K remains though the β-Sn transitions become less sharp and blurred. In fact, the small transitions at 6.48 K is probably explained as superconducting fluctuation effect similar to that of granular tin on graphene reported by Allain et al. and Sun et al.[31,32]. The superconducting fraction of the sample with 267 nm InSb layer is obviously reduced compared to the sample with 100 nm InSb layer. This coincides with the assumption that the strain from the InSb layer stabilizes the α-Sn structure, which is also suggested by the XRD pattern in the right panel of Figure 1(b). Larger thickness of the InSb layer relative to the α-Sn film should give a stronger stabilization effect. The much larger values of resistance of the two samples in Figure 5 compared to that in Figure 4 cannot just be explained by the reduced thickness. It may be related to the more semiconducting character of the Sn films in Figure 5 since the α-Sn is a zero-gap semiconductor while the β-Sn is metallic.

**Summary**
The superconductivity of α-Sn films on two type InSb(001) templates has been investigated. We have confirmed the coexistence of α and β phase with the critical temperature around 3.7 K of β-Sn in the films. Multiple transitions above 4 K with larger critical fields should come from other possible sources such as In contacts, In-Sn alloys or oxides. We eliminated the influence of In contacts and oxides by using Ag paste as contacts and depositing Si protection layer. It is concluded that the In contacts and oxides do not decide the appearance of the superconductivity, though they may induce the aging effects in the film at room temperature. The slow transformation to β phase of the α-Sn films as a consequence of relaxation can still happen under the stabilization of the InSb layer. However, we cannot identified whether the β phase forms during growth or after growth from the present electrical data. We used the InSb/GaSb/GaAs structure to reduce the substrate contribution and investigate the strain effect on the superconductivity by modulating the thickness ratio of the InSb layer to the α-Sn film. We have confirmed the larger strain with the larger ratio can reduce the fraction of the β phase but has less influence on the transition around 5 K. More ratios should be used to examine the strain effect on the film structure and whether it influences the superconductivity through other mechanisms in the future work. The strained α-Sn was predicted to be topologically nontrivial, so the possible topological superconductivity cannot be ruled out. Another possible source of superconductivity is the In atom segregation into the α-Sn film from the InSb layer and formation of In-Sn alloy at the interface during growth. However, we have not observed the sign of this process from transmission electron microscope (TEM) characterization yet.

Further improvement for the quality of the α-Sn films are needed to remove the β phase. The studies on the source of the superconductivity and aging effects and the efforts to make the transitions at higher temperatures complete are still ongoing.

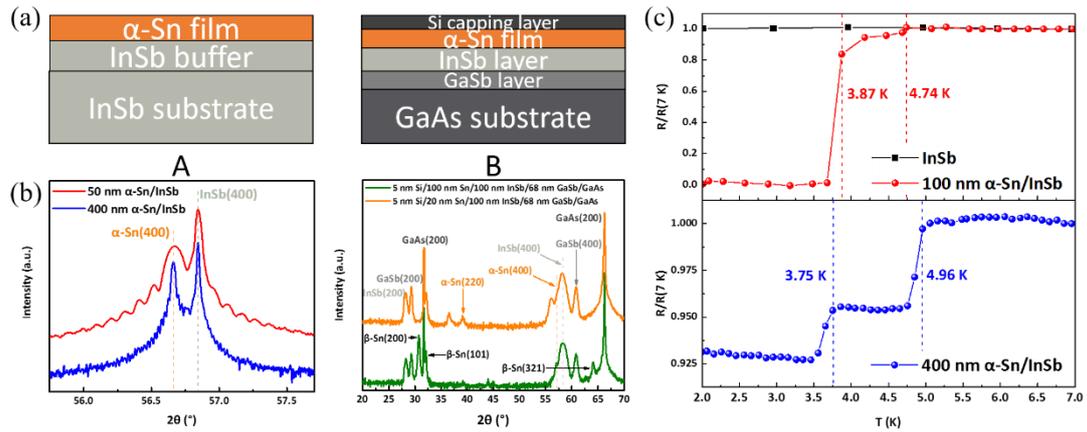

Figure 1. Schematic layer structures and XRD patterns for two type α-Sn samples. (a) Sample structures for two type samples. Type A: α-Sn thin films grown on InSb substrates with InSb buffer. Type B: α-Sn thin films grown on semi-insulating GaAs substrates. The GaSb and InSb layers were grown to give acceptable lattice match with α-Sn. The thickness of the InSb layer was on the order of hundreds nanometers. Some samples were capped with about 5 nm amorphous Si to avoid oxidation. (b) Corresponding XRD patterns for typical samples with structures in (a). Relative diffraction peaks are marked in the figure. (c) Superconductivity for the 100 nm and 400 nm type A samples in (b). The transition temperatures are marked in the figure. The curve for InSb substrate is also demonstrated for reference.

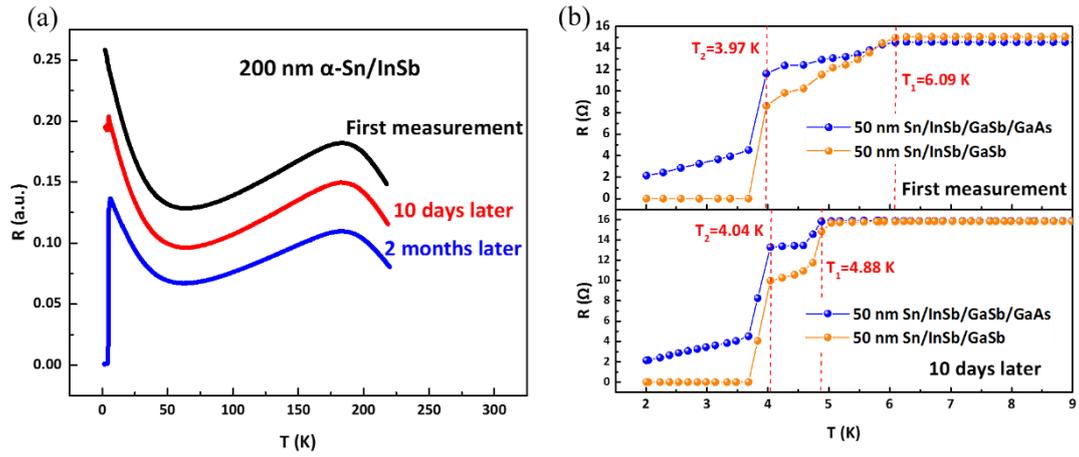

Figure 2. Aging effects on the superconducting properties for the two type α-Sn samples. (a) Resistance-temperature curves of 200 nm type A sample measured repeatedly over time. The curves are offset for clarity. (b) Resistance-temperature curves at low temperature measured with 10 day interval for two type B samples grown on GaAs and GaSb substrates, respectively. The transition temperatures are marked in the figure with red dashed lines.

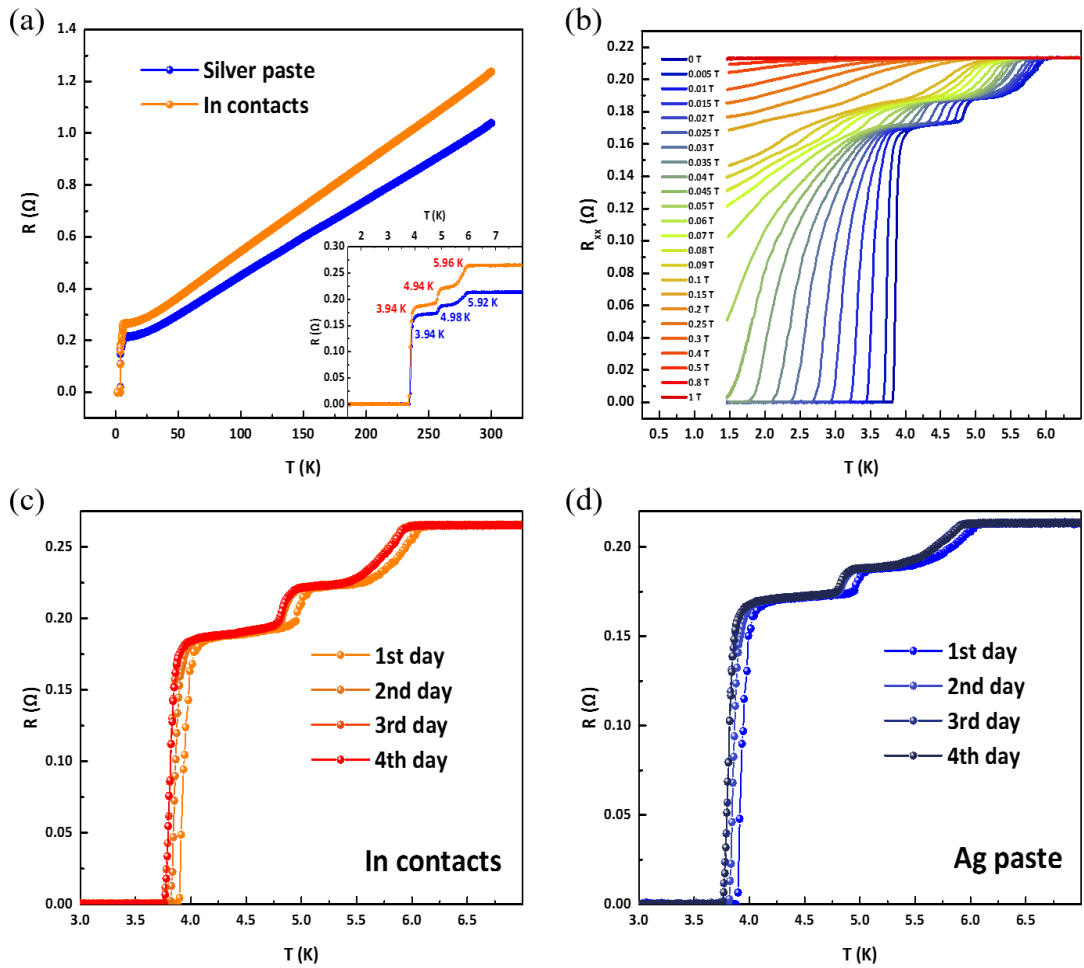

Figure 3. Superconducting properties of the Si capped 100 nm type B α-Sn sample with pressed In and Ag paste on the opposite edges as contacts, respectively. (a) Resistance-temperature curves for pressed In and Ag paste contacts. The inset is the close-up at low temperature showing three superconducting transitions. The corresponding transition temperatures are marked in the inset. (b) Resistance-temperature curves under different magnetic field with Ag paste contacts. (c)(d) Resistance-temperature curves measured over time for In contacts and Ag paste, respectively. The measurement was continuous at low temperature without taking the sample out.

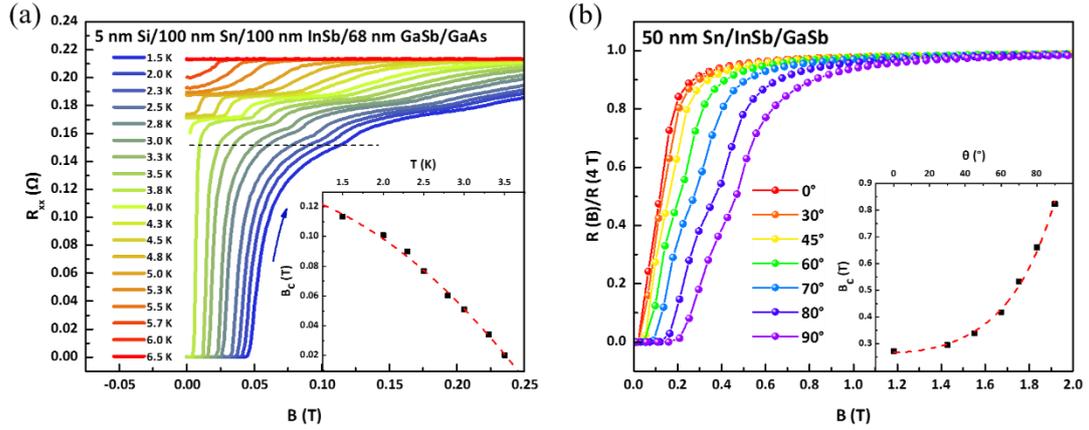

Figure 4. Magnetic field dependence of the resistance for (a) 100 nm type B sample at different temperatures with magnetic field perpendicular to the sample surface. The inset shows the critical fields approximated by setting the constant resistance corresponding to the kinks on the curve at 1.5 K represented by the black dashed line. The red dashed curve is the fitting curve to the experimental data using the equation in the text. (b) 50 nm type B sample at different orientations. 0° and 90° represent the magnetic field perpendicular and parallel to the sample surface, respectively. The inset shows the angular dependence of the critical field fitted with equation (2) in the main text. The critical field is decided by taking the value at which the resistance reaches 90% of the resistance at 4 T.

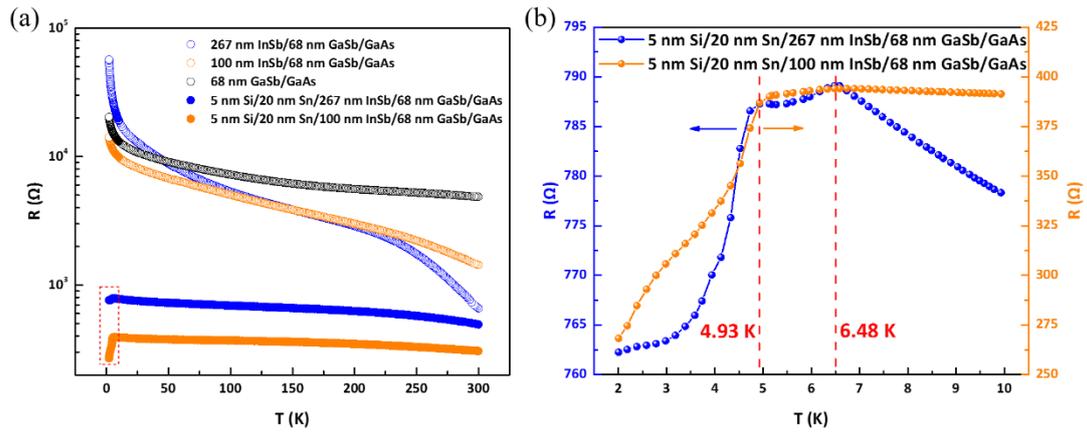

Figure 5. Temperature dependence of the resistance of Si capped 20 nm type B α-Sn samples with different InSb layer thicknesses. Three template samples are shown as reference. (a) Resistance-temperature curves of the gray tin samples and reference templates. (b) The close-up of the resistance-temperature curves of the two 20 nm type B samples in the red dashed rectangle in (a) with different InSb layer thicknesses at low temperature. The corresponding transition temperatures are marked in the figure with red dashed lines.